# Task-specific programming of chaos in neural circuits

Jungyoon Kim[1§], Kyuho Kim[1§], Kunwoo Park[1], Namkyoo Park[2†], and Sunkyu Yu[1*]

[1]Intelligent Wave Systems Laboratory, Department of Electrical and Computer Engineering, Seoul National University, Seoul 08826, Korea

[2]Photonic Systems Laboratory, Department of Electrical and Computer Engineering, Seoul National University, Seoul 08826, Korea

E-mail address for correspondence: [†]nkpark@snu.ac.kr, [*]sunkyu.yu@snu.ac.kr

**Abstract**

Chaotic dynamics have emerged as a versatile resource for neuromorphic and probabilistic computing, enabling high-dimensional nonlinear processing and classical analogues of quantum randomness. Exploiting chaos for computation requires task-dependent control over complexity, as demonstrated in reservoir computing, random-number generation, and probabilistic inference. Existing approaches have focused on tuning element-level parameters, leaving the collective, many-body origin of chaos largely unexplored as a design freedom. Here, we demonstrate programmable chaotic dynamics for task-specific reservoir computing. Using a continuous-time neural-circuit model, we show that tuning network topology drives an ordered-to-chaotic transition, accompanied by transitions in correlation timescales, stability characteristics, and signal propagation. By jointly controlling element-level properties and network topology, we establish a unified chaos-latency phase diagram, revealing that small-world connectivity enables low-latency on-off switching of chaos via edge rewiring. Supported by distinct reservoir-computing benchmarks across various topological regimes, our results demonstrate that network topology serves as a reconfigurable parameter for task-specific computation and tunable randomness.

## Introduction

Unravelling the complexities of chaotic dynamics—characterized by exponential sensitivity to perturbations in the initial state—has been a cornerstone of research across physical, technological, and biological systems, such as network synchronization[1], ray dynamics in deformed microcavities[2], level statistics in quantum many-body systems[3], and neuronal avalanches[4]. Even when each element obeys deterministic laws, the interplay of nonlinearity and feedback can collectively drive the system into a chaotic regime, as opposed to ordered dynamics such as quiescent states[5], oscillation-quenched states[6,7], and regular oscillations[5,8]. Motivated by the rich chaotic activities observed in biological neural circuits[9-12] and by the essential role of randomness in quantum devices[13,14], recent efforts have sought to harness chaos for computation, including reservoir computing near the edge of chaos[15-18] and probabilistic computing that leverages chaos to emulate quantum randomness[19].

For such computing paradigms, a central challenge is the realization of programmable chaos: achieving energy-efficient and reconfigurable transitions between ordered and chaotic dynamics. A representative approach is to control individual elements, where order-chaos transitions are induced by tuning a subset of element-level parameters, such as the effective neuronal gain or coupling strength in random networks[9], the feedback gain in delay-based reservoirs[20], and the bias conditions in oscillators[21]. Despite their distinct configurations, these approaches share a common operating principle: identifying the boundary between order and chaos to determine the optimal control point.

From the perspective that chaotic dynamics arise from interactions among many elements, a complementary control axis to element-level engineering is the manipulation of network topology. This degree of freedom has been explored in relation to noise immunity in signal processing[22-24] and system controllability[25,26]. Even with fixed elemental dynamics, altering network topology can manipulate paths, clustering, and feedback loops, which shape signal spreading and memory functions[27,28]. Moreover, recent work has further emphasized that network topology can affect collective dynamics in complex and excitatory-inhibitory networks[28,29]. However, although chaotic dynamics in random networks[9,12,30] and computation utilizing small-world properties[31,32] have been studied separately, a unified framework linking element-level dynamics, topology-controlled chaos, and task-dependent computational performance has not yet been established.

Here, we establish a task-specific framework for configuring neural circuits for reservoir computing, enabled by programmable chaotic dynamics through the coordinated control of element-level properties and network topology. We show that tailoring network topology induces a transition from ordered to chaotic dynamics, characterized by correlation timescales, Lyapunov stability, and signal propagation. This observation identifies a phase transition across small-world topology, allowing for the rapid and reconfigurable switching of chaos via network rewiring. By examining the element-level and network-level parameter spaces to construct a chaos-latency phase diagram, we demonstrate that distinct dynamical regimes are selectively suited to different tasks—with small-world networks near the edge of chaos being particularly favourable for tasks simultaneously requiring low latency and enhanced retention—thereby establishing a link between topology-controlled chaos and task-specific functionality. Our results provide the essential features for probabilistic computing, reservoir learning, and random number generation, while establishing a novel classification of complex networks based on chaos and latency.

## Results

### Model definition

We consider a continuous-time neural circuit model (Fig. 1a) introduced in the context of spin-glass-inspired neural dynamics[9]. For an $N$-node circuit, the governing equation is

$$\frac{dx_i(t)}{dt} = -x_i(t) + \sum_{j=1}^{N} W_{ij}\varphi_j(t) , \quad (1)$$

where $x_i(t)$ and $\varphi_i(t)$ denote the time-dependent membrane potential and the corresponding nonlinear firing rate of the $i$-th neuron, respectively, and $W_{ij}$ denotes the $(i,j)$ entry of the weight matrix $W$. The neuronal activation is fixed as $\varphi_i(t) = \tanh(x_i(t))$, which captures the bounded, odd, and nonlinear nature of neuronal responses[9]. Motivated by Dale's law[33,34] for biological neurons, the circuit consists of excitatory and inhibitory neurons, whose cell types determine the signs of their outgoing synapses: $W_{ij} = +1$ and $W_{ij} = -1$ for the $j$-th excitatory and inhibitory neurons (Fig. 1b), respectively (see Methods). For a given excitatory fraction $E \in [0,1]$, the inhibitory fraction is $1 - E$, and the cell type of each neuron is independently assigned according to a Bernoulli distribution with parameter $E$.

In the proposed neural circuit, the network topology is fully encoded in the weight matrix $W$. A prior study[9] has extensively investigated the neuronal dynamics when $W$ is modelled as an unstructured random matrix with independent Gaussian entries, as in the Ginibre orthogonal ensemble[35]. Extensions incorporating excitation-inhibition balance[10], nonlinear transfer-function effects[12], and coupled neuronal-synaptic dynamics[36] have further characterized transitions to chaos in random neural networks. Beyond such dense random-matrix models, sparse networks have also been examined to explore the transition from ordered to chaotic dynamics through spectral engineering of the weight matrix[37]. However, these sparse networks are based on random graphs close to the Erdős-Rényi model[38] with strongly suppressed clustering and short path lengths. Therefore, how tailored network topology induces the transition from order to chaos remains elusive.

We extend this random-network model to a range of network topologies using the following procedure suggested in the previous study[39]. From the normalized distance, $d_{ij} = D_{ij}/D_{\max}$, where $D_{ij} = \min(|i-j|, N-|i-j|)$ and $D_{\max} = \lfloor N/2 \rfloor$, the probability that an edge exists between the $i$-th and $j$-th nodes is given by $p_{ij} = \beta p_0 + (1-\beta)\Theta(p_0 - d_{ij})$, where $\beta \in [0,1]$ denotes the disorder parameter, $\Theta$ is the Heaviside step function, and $p_0$ is the parameter determining the sparsity of $W$. Because $\beta$ corresponds to the rewiring probability in the Watts-Strogatz formalism[22], the resulting circuit spans a broad range of network topologies from a regular graph ($\beta \to 0$), through a small-world graph, to an Erdős-Rényi graph ($\beta \to 1$)[38]. By solving Eq. (1) for different realizations of $W$ (see Methods), we characterize the circuit dynamics (Figs. 1c-1e), which show distinct transitions between ordered and chaotic behaviour depending on the underlying network topology.

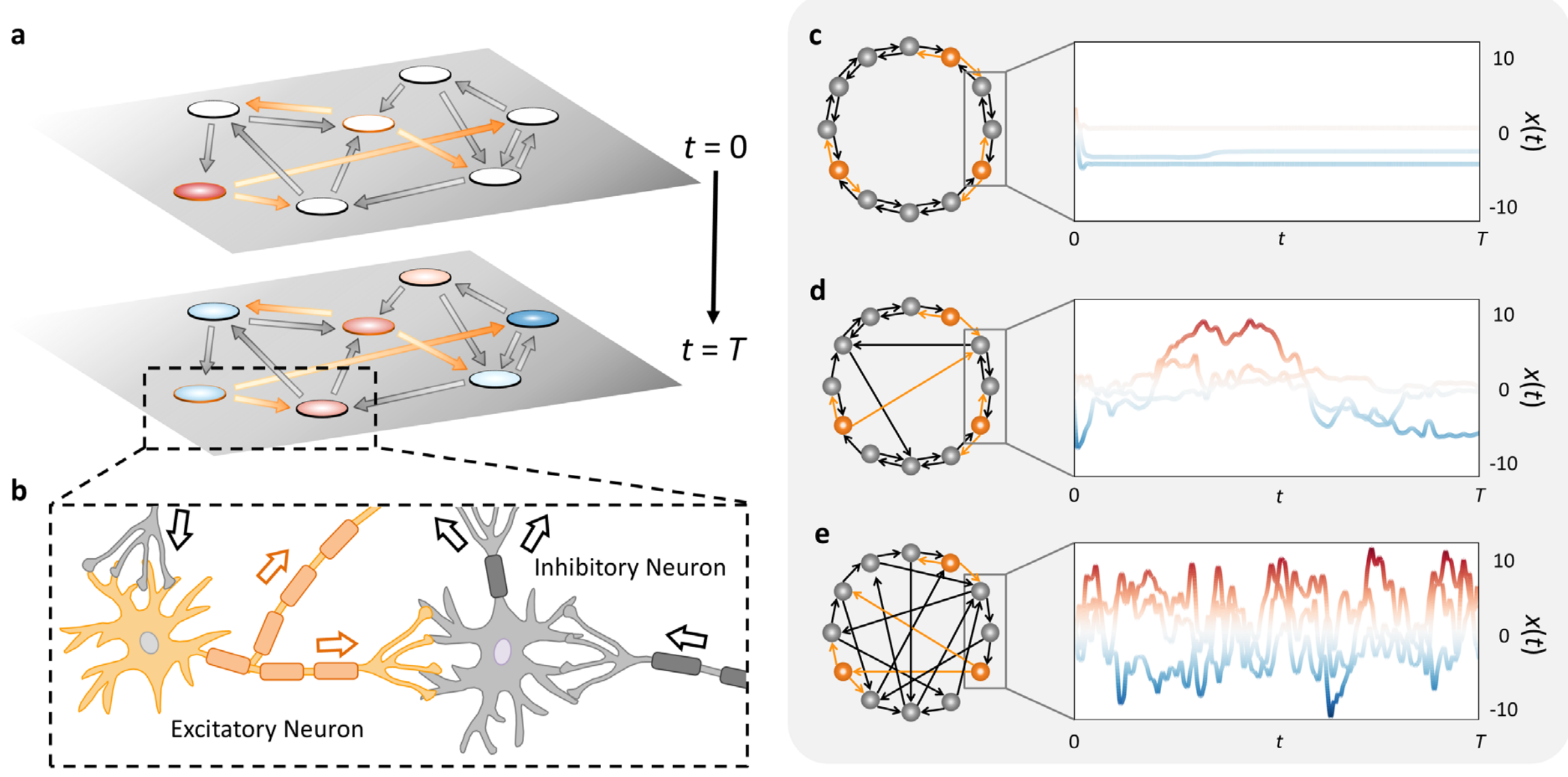

**Figure 1. Network topology on neuronal dynamics. a,** A directed neural circuit network with node-wise positive (orange) or negative (black) edges. The initial values are uniformly sampled from the interval [–5, 5]. The node states evolve dynamically over $t$. **b,** Illustration of Dale-type sign constraints, where individual neurons provide strictly excitatory or inhibitory connections. **c-e,** Dynamics in the regular (**c**) small-world (**d**), and random (**e**) networks for $\beta = 10^{-4}$, 0.1, and 0.8, respectively. $N = 300$, $p_0 = 0.07$, and $E = 0.2$ in **c-e**.

**Topology-dependent dynamics**

To quantify how network topology governs circuit dynamics, we introduce two distinct temporal metrics: the memory timescale[36] $\tau_c$ and the propagation latency $\tau_p$, which are defined as

$$\tau_c = \int_0^\infty d\zeta \left[ \frac{Q(\zeta)}{Q(0)} \right]^2, \; \tau_p = \left\langle \frac{1}{N - |B|} \sum_{i \notin B} \min_t (|x_i(t)| \geq \epsilon) \right\rangle_B, \tag{2}$$

where $Q(\varsigma) = \langle \varphi_i(t)\varphi_i(t+\varsigma) \rangle_{i,t}$ is the autocorrelation of the dynamical system, $\varepsilon$ is a predefined signal threshold, and $B \subset \{1, 2, \ldots, N\}$ represents a subset of initially perturbed nodes (see Methods). Notably, $\tau_c$ captures the duration over which the system preserves memory of its prior states, serving as a metric for characterizing chaotic dynamics: a smaller $\tau_c$ indicates more randomized, memoryless dynamics. On the other hand, $\tau_p$ measures the average first-passage time for the propagation of initial perturbations, characterizing the latency of global activation[27]. Consequently, we can evaluate the chaos degree and speed of circuit dynamics using $\tau_c$ and $\tau_p$.

We evaluate these temporal metrics in relation to the static network topology. Figures 2a and 2b show the variations in network parameters and temporal metrics, respectively, as functions of the disorder parameter $\beta$ with a fixed excitatory fraction $E = 0.2$ (Supplementary Note S1 for the robustness of $\tau_p$ over the hyperparameter $\varepsilon$). We employ the clustering coefficient $C$ and the average path length $L$ (see Methods) as static descriptors for network topology, which conventionally characterize the transition from regular to random networks through the small-world regime as $\beta$ increases[22]. In our model, these static quantities directly correlate with the circuit dynamics. First, the high clustering coefficient $C$ at low $\beta$ reflects the abundance of local feedback loops across the network. Such loops enable the slow decay of the temporal correlation $Q(\varsigma)$ via recurrent dynamics (Fig. 2c), eventually enhancing the memory performance characterized by $\tau_c$. On the other hand, the path length governs the propagation latency of the circuit dynamics, as evidenced by the approximately linear spreading of perturbations relative to the graph distance— that is, the number of edges along the shortest path—to the initially perturbed nodes in regular networks and the near-instantaneous activation achieved with the emergence of shortcuts (Fig. 2d).

The analysis reveals that the classification of static network topology parallels the dynamical classification of neural circuits in terms of chaos and latency. In the regular network limit, dynamics are stable and propagate slowly, whereas random networks manifest highly chaotic and rapidly spreading dynamics. The unique intermediate regime—the small-world network—is characterized by near-critical, regular dynamics with significantly suppressed latency. Such a framework suggests that each topological phase serves distinct functional roles when utilized as a computing platform.

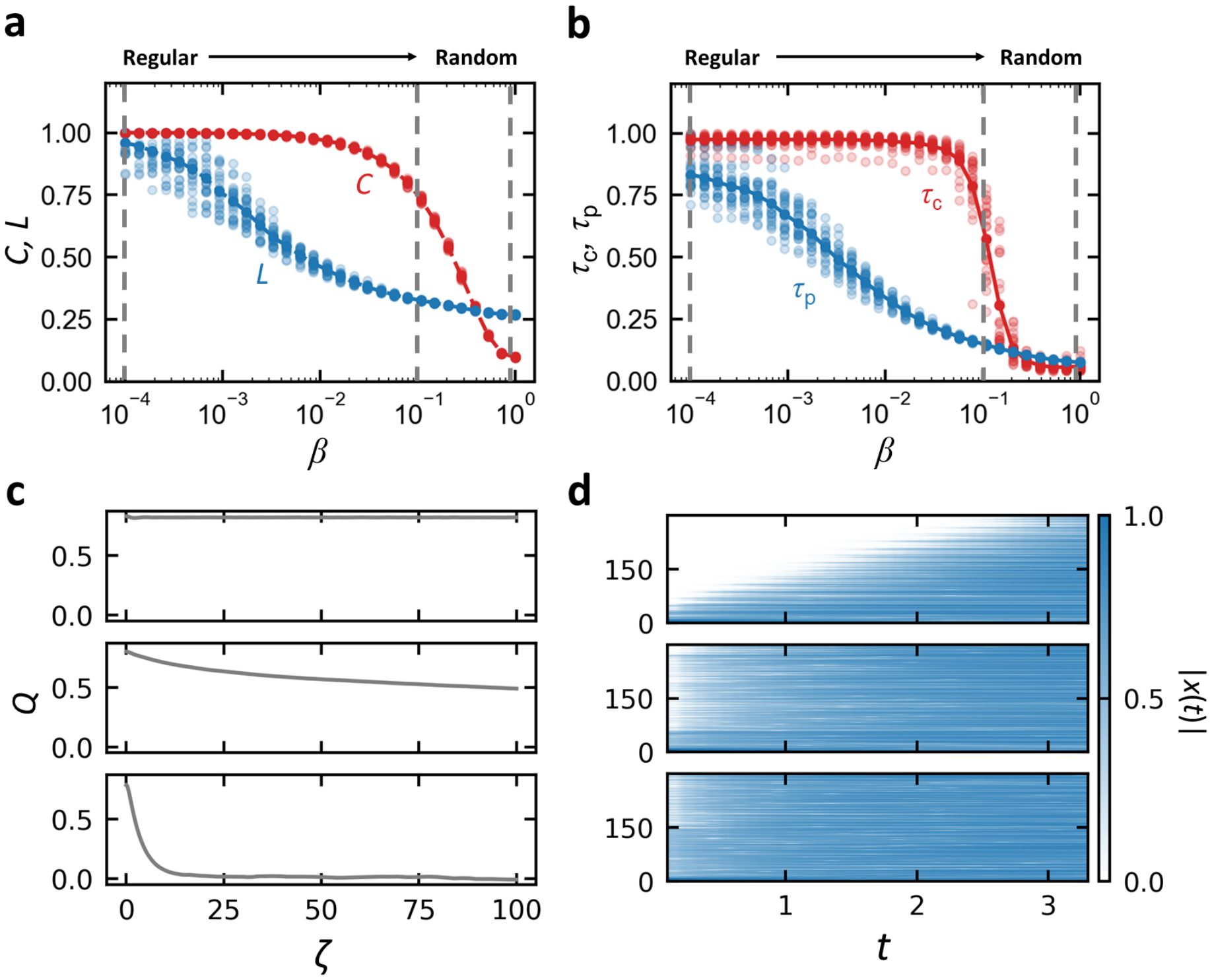

**Figure 2. Topological classification of circuit dynamics. a,** Network parameters: clustering coefficient $C$ and average path length $L$, as functions of $\beta$. **b,** Temporal metrics: memory timescale $\tau_c$ and first-passage time $\tau_p$, as functions of $\beta$. $\varepsilon = 0.01$ in estimating $\tau_p$. **c,d,** Time autocorrelation $Q(\varsigma)$ (**c**) and perturbation spreading (**d**) for regular (top; $\beta = 10^{-4}$), small-world (middle; $\beta = 0.1$),

and random (bottom; $\beta = 0.8$) networks. The $y$-axis of each subpanel in **d** depicts nodes sorted by their shortest-path graphical distance from the batch of initially perturbed nodes $B$. Dashed lines in **a** and **b** denote $\beta$ values for the subpanels in **c** and **d**. In analysing dynamics, we generate an ensemble of 24 random realizations for each $\beta$ value. All metrics in **a** and **b** are normalized by their maximum values. All other parameters are the same as those in Figs. 1c-1e.

**Phase diagrams**

We extend our topology-based classification by incorporating element-level characteristics—the excitatory fraction $E$—thereby establishing a comprehensive phase diagram over the ($\beta$, $E$) parameter space (Fig. 3). To characterize the resulting chaotic dynamics across this space, we again employ the memory ($\tau_c$; Fig. 3a) and latency ($\tau_p$; Fig. 3b) metrics, which reveal distinct dynamical regimes governed by the excitatory fraction (Supplementary Note S2 for size dependency). At high $E$ value ($0.55 \lesssim E \lesssim 1.00$), representing excitatory-dominant regimes, the system maintains ordered dynamics with long-term memory that remains largely independent of the network topology. In contrast, at intermediate $E$ ($0.35 \lesssim E \lesssim 0.55$), the system exhibits chaotic dynamics with memoryless responses regardless of the topological configuration. In the remaining regimes, the transition between ordered and chaotic behaviours can be actively modulated by altering the network topology. Consequently, the synergistic tailoring of element-level features ($E$) and network topology ($\beta$) enables the precise programming of chaotic dynamics (arrows in Fig. 3a).

As a single-parameter descriptor of the observed dynamics, we employ the largest Lyapunov exponent (LLE), $\lambda_{max}$, of the system: $\lambda_{max} > 0$ indicates chaotic dynamics, $\lambda_{max} < 0$ corresponds to stable dynamics, and $\lambda_{max} = 0$ to marginally stable—that is, fixed-point or oscillatory—behaviours[40] (see Methods). Figure 3c shows $\lambda_{max}$ over the parameter space ($\beta$, $E$). We note that the chaotic phase characterized by $\lambda_{max} > 0$ (blue region in Fig. 3c) coincides well with the memoryless phase (white region in Fig. 3a) where $\tau_c$ is small. Within this chaotic regime, an increase in $\lambda_{max}$ (deeper blue region in Fig. 3c) indicates faster divergence of trajectories as $\beta$ increases, consistent with the reduction in latency (Fig. 3b). It is worth noting that the LLE further subdivides the non-chaotic phase into marginally stable ($\lambda_{max} \approx 0$) and quiescent ($\lambda_{max} < 0$) regimes, thereby providing a more refined classification of the ordered dynamics.

Based on the phase diagram, we demonstrate reconfigurable transition between regular to chaotic dynamics using network rewiring: a directed edge swap, referred to as the Maslov-Sneppen rewiring[41]: select two separate edges ($u \rightarrow v$) and ($t \rightarrow s$), and then, swap the edges so that the resulting edges are ($u \rightarrow s$) and ($t \rightarrow v$) (Fig. 3d). This increases network randomness while preserving all the indegrees and outdegrees of the network. Notably, for an initial point near the critical $\beta$ value, swapping only 6% of edges drives the transition from a stable, regular state to a chaotic state (Fig. 3e), highlighting the efficient programming of chaos near the phase boundary.

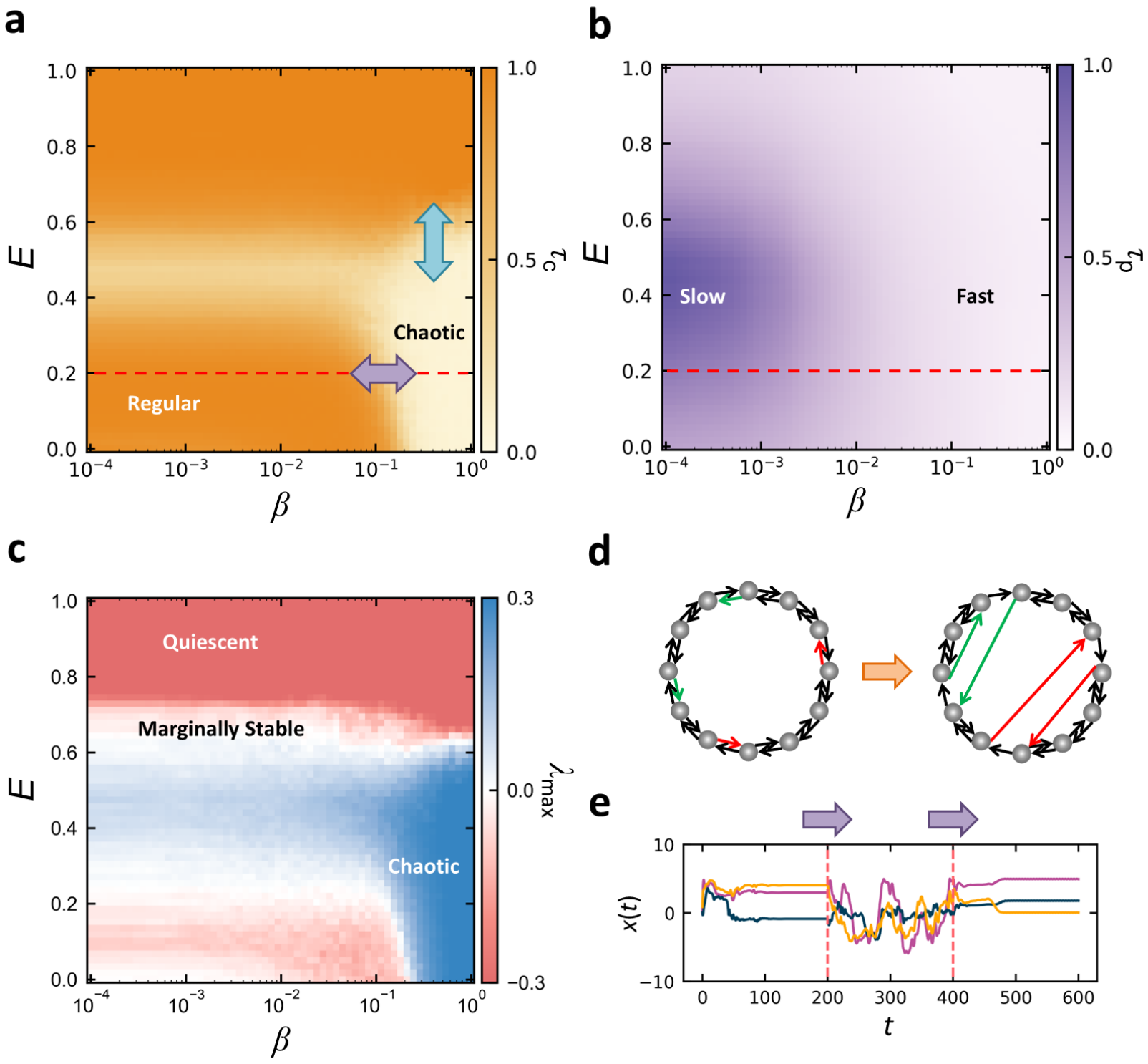

**Figure 3. Chaos-latency phase diagrams**. Phase diagrams are shown over the ($\beta$, $E$) parameter space for the memory metric $\tau_c$ (**a**), latency metric $\tau_p$ (**b**), and the LLE (**c**). The parameters $\beta$ and $E$

values are swept over the ranges [$10^{-4}$, $10^{0}$] and [0.0, 1.0], respectively, with a resolution of 60 × 60. Each point in the parameter space represents an average over an ensemble of 24 realizations. Red dashed lines in **a** and **b** denote $E = 0.2$. **d,** Schematic for edge rewiring. The green and red edges indicate the selected rewired edge pairs. **e,** Phase transition of dynamics induced by edge rewiring, which corresponds to the purple arrow in **a**: starting from $\beta = 10^{-1.15}$ (total 6015 edges) and performing 180 swaps at $t = 200$, then returning to the original state at $t = 400$. All other parameters are the same as those in Fig. 2.

**Learning with programmed chaos**

Diverse computational tasks impose distinct requirements on dynamical complexity, necessitating different degrees of chaos, as demonstrated in real-time computation near the edge of chaos[42], input sequence memory using locally expansive stable dynamics[43], suppression of chaos using predictive alignment[44], and analogue computing leveraging chaotic attractors[45]. Consequently, the rich dynamical phases supported by our neural circuit—enabled through comprehensive engineering of network topology and element-level control—motivate its use as a universal framework for a broad class of tasks. As an example, we apply our neural circuit, within a reservoir-computing framework, to three distinct tasks (Fig. 4a): image classification (Fig. 4b), remote signal tracking (Fig. 4c), and temporal signal prediction (Fig. 4d).

We utilize our circuit to construct an echo state network (ESN) architecture[46] (Fig. 4a). To program chaos, we focus on tailoring the network topology while keeping the excitatory fraction fixed at $E = 0.2$ (red dashed lines in Figs. 3a and 3b). By varying the topological parameter $\beta$, the neural circuit undergoes three distinct dynamical regimes based on $\tau_c$ and $\tau_p$: high-latency ordered (HO) dynamics ($\beta \lesssim 10^{-2}$), low-latency edge-of-chaos (LE) dynamics ($10^{-2} \lesssim \beta \lesssim 10^{-0.8}$), and low-latency chaotic (LC) dynamics ($\beta \gtrsim 10^{-0.8}$). From a network-topology perspective, these regimes correspond directly to regular, small-world, and random networks, with the small-world regime operating at the edge of chaos (see Methods). Together with memory ($\tau_c$) and latency ($\tau_p$) characteristics, the network topology determines the expressivity of the ESN architecture (see Supplementary Note S3), thereby governing the nonlinear representation capacity and trainability of the network[47,48].

First, we evaluate the reservoir performance using image classification on the MNIST dataset[49], which is formulated as a time-varying signal flow task. We devise a time-series learning architecture as follows: static images are presented to the network for a finite duration, allowing the reservoir states to evolve toward an input-dependent representation, and a linear readout is trained on the resulting reservoir states to perform the final classification. This architecture can be understood as a reservoir-based temporal embedding of static inputs, in which spatial image patterns are converted into transient network responses and subsequently mapped onto class labels. Figure 4b shows that this classification task, possessing strong correlation across dataset, is well-suited to regular networks yielding HO dynamics ($\beta \lesssim 10^{-2}$), which allow for long-term memory.

As an example of a task well suited to the opposite extreme of network topology—random networks—we consider chaotic temporal-signal prediction using the Lorenz-96 signal[50], one of the standard benchmarks for evaluating reservoir computing systems. We evaluate forecasting performance at prediction horizons of $\Delta t = 1$ and $\Delta t = 5$. As shown in Fig. 4d, chaotic networks with LC dynamics ($\beta \gtrsim 10^{-0.8}$) consistently outperform regular and small-world networks at both prediction horizons. This superiority highlights the computational advantage of strong nonlinear expansion and sensitivity to input variations—intrinsic features of chaos—for hardware systems tasked with accurately forecasting chaotic dynamics.

Inspired by prior studies on memory capacity in small-world networks[32,51] and signal tracking tasks[52], we test our neural circuit on another challenging task: tracking a random-valued, piecewise-constant input stream signal at a remote readout. In this task, the input signal is injected into a single node and decoded from a spatially distant readout block using only the reservoir state sampled at the end of each input period. This configuration requires two dynamical capabilities: rapid propagation of the input signal from the injection node to the remote readout block within a single period—that is, low latency—and sufficient retention of the transmitted information to recover the current drive value before the period expires—that is, strong memory retention. For this challenging task, small-world networks with LE dynamics ($10^{-2} \lesssim \beta \lesssim 10^{-0.8}$) achieve the best performance (Fig. 4c), owing to their low latency and favourable balance between memory capacity and nonlinear representation. Remarkably, the performance profile across $\beta$ provides a clear signature of edge-of-chaos operation: increasing chaos initially improves performance by enhancing dynamical richness and separability along with reduced low latency, whereas further progression into the strongly chaotic regime rapidly degrades performance because excessive instability suppresses memory retention.

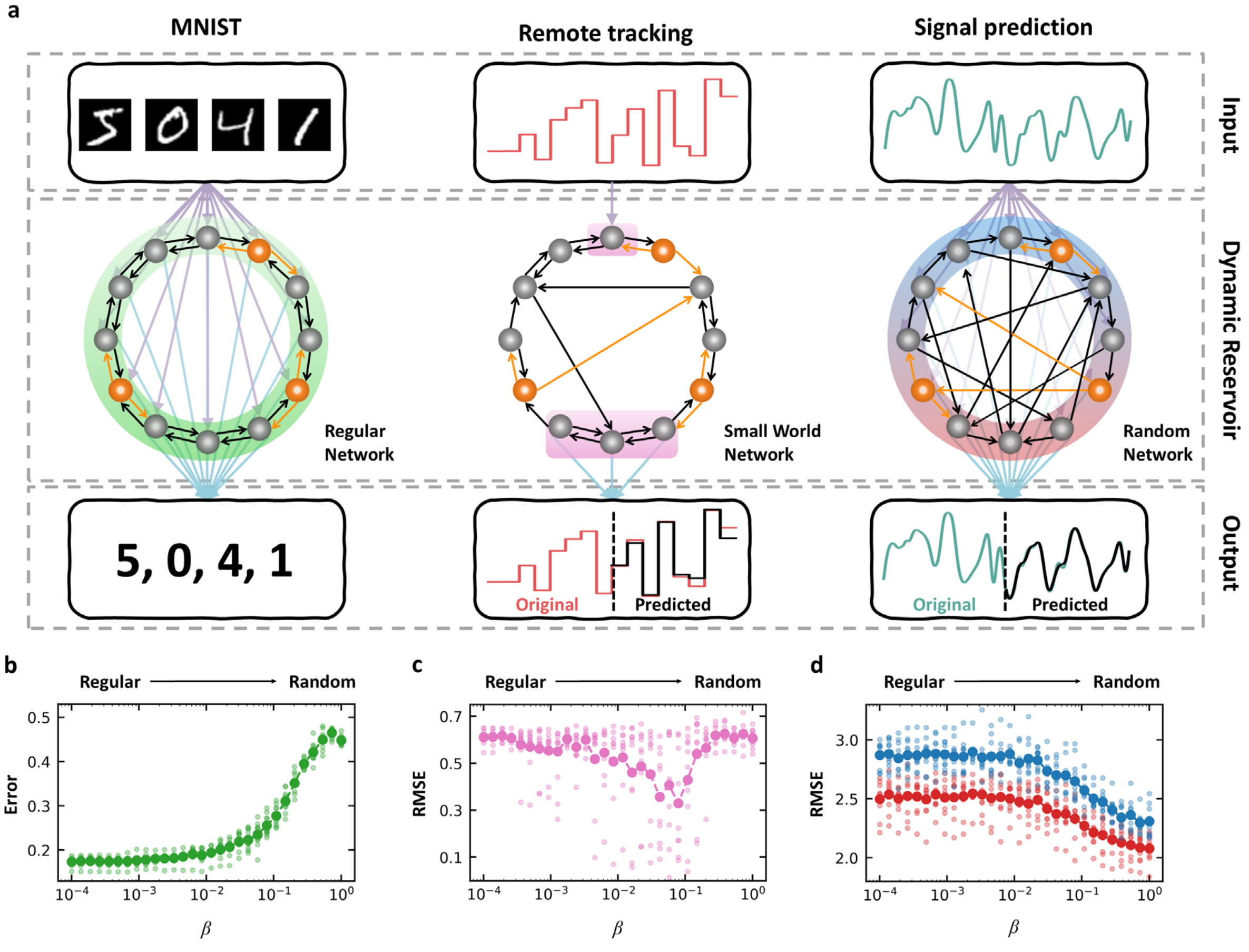

**Figure 4. Topologically programmable chaos for task-specific computation**. **a,** Schematic for reservoir-computing framework based on our neural circuit, applied to three representative tasks: MNIST image classification (left), remote signal tracking (centre), and temporal signal prediction (right). **b-d,** Task performances as a function of $\beta$ at $E = 0.2$, spanning regular, small-world, and random network regimes: classification error on MNIST (**b**), root-mean-square error (RMSE) for remote signal tracking (**c**), and RMSE for chaotic time-series prediction at $\Delta t = 1$ (red) and $\Delta t = 5$ (blue) (**d**). Each point in **b-d** denotes one random realization of a neural circuit.

## Discussion

As shown in the examples of programmable chaos for diverse computing tasks, our neural circuit serves as a fundamental framework for exploiting transitions between ordered and chaotic dynamics. By systematically varying the disorder parameter $\beta$ and excitatory fraction $E$, the memory timescales, signal propagation latency, and expressivity of the system can be comprehensively engineered. By integrating these topological and elemental parameters directly into the learning loop, future neural circuits could autonomously reconfigure their connectivity and excitatory–inhibitory balance in real time[53]. Such an extension would enable a programmable neuromorphic hardware to shift dynamically between computational modes—transitioning from stable memory states to expansive chaotic regimes—thereby achieving high-dimensional, reconfigurable, and stable learning frameworks for diverse challenging tasks.

Our findings extend prior studies of chaos in neural circuits, which have largely focused on random network models that incorporate element-level control parameters, such as synaptic gain[9] or plasticity[36]. In such settings, the network is typically assumed to lack significant short-range structures, leading to suppressed clustering and localized feedback loops. Under this assumption, temporal correlations are well approximated by a single-node stochastic process within the dynamical mean-field theory (DMFT) framework[9,12,36]. By contrast, our results show that structured topology—especially small-world connectivity—provides a control axis for dynamical regimes[29]. This demonstrates that the order-chaos transition is governed not only by element-level parameters or global coupling strength, but also by mesoscale network structures.

Our results motivate the novel design principle in the hardware implementation of neuromorphic reservoir computing. Memristor-CMOS routing fabrics[54], manycore neuromorphic processors[55] and scalable neurosynaptic chips[56] could realize topology-tunable circuits in which connectivity patterns and excitation-inhibition structure are simultaneously reconfigured to select task-dependent dynamical regimes. In photonic platforms, saturable gain and loss nonlinearities can provide neuron-like dynamics[57], couplings between optical neurons can be engineered through integrated photonic networks[58] or scattering networks[59]. In such hardware platforms, engineering network complexity in relation to chaotic dynamics may open a route toward adaptive reservoir computing, low-power probabilistic computation, and reconfigurable neuromorphic systems.

In conclusion, we demonstrated programmable chaos in Dale-constrained dynamic neural circuits by jointly tuning network-topological randomness and the excitatory fraction, thereby constructing a unified chaos-latency phase diagram. By engineering the network topology, the system can be driven among stable, edge-of-chaos, and strongly chaotic regimes with tunable memory, propagation speed, and expressivity. This neural-circuit framework provides a universal, programmable learning architecture for a broad range of computing tasks.

## Methods

**Network configuration.** The $j$-th neuron is assigned a cell type $\sigma_j \in \{+1, -1\}$, sampled independently and identically from a Bernoulli distribution with parameter $E$. Given a directed adjacency matrix $A$ with $A_{ij} \in \{0, 1\}$, the synaptic weight matrix is defined as $W_{ij} = A_{ij}\sigma_j$. Prior studies of such neural networks typically use the scaling $W_{ij} \sim N^{-1/2}$, which yields an input scale of order $N^{1/2}$. Our binary coupling choice $W_{ij} = \pm 1$ can be related to this convention as follows. In a sparse network with edge density $p_0$, the typical absolute input scale is of order $Np_0$. For the small-world generation algorithm used here, the in- and out-degrees are sums of independent Bernoulli variables and therefore do not exhibit heavy-tailed statistics. Thus, choosing $p_0 \sim N^{-1/2}$ yields an input scale comparable to that in prior studies. In our setting, we fix $p_0 = 0.07$, which keeps the overall input magnitude within a numerically stable range for the finite system sizes considered.

**Dynamics solver.** We solve Eq. (1) using the fifth-order Dormand–Prince method with adaptive step size, implemented in the JAX-based Python library Diffrax[60]. Unless otherwise specified, simulations are run up to $t_{\max} = 300$ with a sampling interval of $\Delta t = 0.1$, and asymptotic quantities are estimated from these finite-time trajectories. For all variables, an initial washout period of $t_{\text{washout}} = 10$ is discarded. For each point in the $(\beta, E)$ parameter space, we examine 24 independent realizations and use the ensemble average to characterize the dynamics.

**Details in measuring dynamics.** To quantify perturbation spreading using $\tau_p$, we initialize the network, select a subset $B \subset \{1, 2, \ldots, N\}$ consisting of a contiguous block of nodes, and perturb nodes in $B$. Because the underlying small-world construction is based on a one-dimensional ring lattice, proximity in index space corresponds directly to proximity in graph distance in the absence of rewiring. Therefore, this procedure ensures that perturbations are spatially localized on the underlying graph. We select $|B| = 10$ and $\varepsilon = 0.01$, and average over the choice of $B$ (Supplementary Note S1 for robustness to $\varepsilon$).

**Network metrics.** We calculate the network clustering and average shortest path length using the unweighted version of the network, that is, without considering the excitation-inhibition weights, while considering directedness. Given a directed adjacency matrix $A$, the clustering coefficient for the $i$-th node is defined as

$$C_i(A) = \frac{(A + A^{\top})^3_{ii}}{2[d_i^{\text{tot}}(d_i^{\text{tot}} - 1) - 2d_i^{\leftrightarrow}]}, \tag{3}$$

where $d_i^{\text{tot}} = \Sigma_{j\neq i}(A_{ij} + A_{ji})$ is the total degree of node $i$ and $d_i^{\leftrightarrow} = \Sigma_{j\neq i}(A_{ij}A_{ji})$ is the number of bilateral edges between $i$ and its neighbours[61]. The clustering coefficient is defined as the average of $C_i(A)$ over all nodes $i$. The average shortest path length is defined as the mean over shortest paths between all possible node pairs, where each shortest path length is calculated using Dijkstra's algorithm. We calculate all parameters using the NetworkX library[62].

**Lyapunov analysis.** The LLE is defined as the asymptotic exponential growth rate of the magnitude of an infinitesimal perturbation vector $v(t)$[40]:

$$\lambda_{\max} = \lim_{t\to\infty} \frac{1}{t} \log \frac{\|v(t)\|}{\|v(0)\|}. \tag{4}$$

In evaluating the LLE, we utilize the explicit form of the Jacobian from Eq. (1):

$$J = -I_N + W \,\mathrm{diag}(\mathrm{sech}^2(x)), \tag{5}$$

where $I_N$ is the $N$-dimensional identity matrix, and diag($d$) denotes the diagonal matrix with the diagonal components of the vector $d$. Therefore, we can evolve $v(t)$ with the linearized state trajectory, as follows:

$$\frac{dv_i(t)}{dt} = -v_i(t) + \sum_{j=1}^{N} W_{ij} \,\mathrm{sech}^2(x_j(t)) v_j(t). \tag{6}$$

The LLE is estimated from the time-averaged logarithmic growth rate of the perturbation norm up to time $t_{\max} = 2000$.

**Reservoir computing tasks.** We implement a reservoir computing framework in which the network dynamics are driven by external inputs, and inference is performed using a linear readout trained on the resulting reservoir states. For a given network with weight matrix $W$, the input-driven dynamics follow

$$\frac{dx_i(t)}{dt} = -x_i(t) + \sum_{j=1}^{N} W_{ij} \phi_j(t) + s_i(t), \tag{7}$$

where $s_i(t)$ is the $i$-th component of the input signal. For each task, reservoir states $x_i(t)$ of pre-selected output nodes are collected to form an $M \times N_{\text{out}}$ feature matrix $Z$, where $M$ is the number of training samples and $N_{\text{out}}$ is the number of output nodes. The corresponding target outputs are arranged in $Y$, a dimension $M \times d_{\text{target}}$ matrix, where $d_{\text{target}}$ is the dimension of the target. We only train the output linear layer $W_{\text{out}}$ by solving the regularized least-squares problem

$$W_{\text{out}} = \arg\min_{W} \left\{ \frac{1}{M} \|ZW - Y\|_{\mathrm{F}}^2 + \alpha \|W\|_{\mathrm{F}}^2 \right\}, \tag{8}$$

where $\alpha$ is a regularization parameter chosen to be $\alpha = 10^{-4}$ and $\|\cdot\|_F$ denotes the Frobenius norm. To ensure statistical reliability, performance for each task is evaluated over 10 random realizations of the neural circuits for each value of $\beta$.

In the MNIST classification task, we rescale the original images to 20 × 20 and flatten them to match the reservoir size $N = 400$. This input is scaled by $N^{-1/2}$, injected into the reservoir directly as a static input term and is presented for a duration of $t_{\text{input}} = 10$. The output weights are trained using the full reservoir state. Performance in Fig. 4b is measured by the error rate, defined as the fraction of test images incorrectly classified by the reservoir.

In the remote signal tracking task, we generate uniformly random target values, $u_k \sim \text{Unif}[-1, 1]$ ($k = 0, 1, 2, \ldots$). For a given period value $T = 5$, the scalar input time series is defined as $s(t) = \Sigma_k u_k \mathbf{1}_{[kT, (k+1)T)}(t)$ during $0 \le t \le t_{\max} = 2000$, where the indicator function satisfies $\mathbf{1}_A(t \in A) = 1$ and $\mathbf{1}_A(t \notin A) = 0$. This definition comprises a piecewise-constant stream of randomly selected target values (Fig. 4a). The signal is injected into a single node, scaled by a predefined amplitude value of 0.5. Using the reservoir states measured from a spatially remote block of 30 nodes located at lattice distance $N/2$ from the input node, a linear readout is trained to estimate the current signal $u_k$ from the firing rate features sampled at the deadline $t = (k + 1)T - 10^{-9}$, immediately before the input is updated to the next symbol. After discarding an initial washout of 50 symbols, the remaining samples are split into training and test sets with a training ratio of 0.7. Performance in Fig. 4c is quantified by the test RMSE.

In the chaotic-signal prediction task, we use the Lorenz-96 system, whose state trajectory $r(t) = [r_0, \ldots, r_{d-1}]^{\mathrm{T}}$ evolves according to the dynamical equation[63]:

$$\frac{dr_i}{dt} = (r_{i+1} - r_{i-2})r_{i-1} - r_i + F \tag{9}$$

for $i = 0, \ldots, d-1$, where we assume periodic boundary conditions $r_d = r_0$, $r_{-1} = r_{d-1}$ and $r_{-2} = r_{d-2}$. In our example, we choose the standard forcing value $F = 8$, and set the system dimension $d$ to 10. The Lorenz-96 trajectory $r(t)$ is embedded into the input vector as $s(t) = Ur(t)$, where $U$ is an $N \times 10$ linear projection matrix whose entries are independently sampled from a mean zero Gaussian distribution with standard deviation 0.1. The reservoir is driven by the full Lorenz-96 trajectory, while the prediction target is chosen as a single component, $r_0(t)$. The reservoir dynamics are simulated over a duration of $t_{\max} = 200$ with a time step of $dt = 0.1$, and an initial washout period of $t_{\text{washout}} = 20$ is discarded. After the washout, the reservoir states $x(t)$ are paired with the future target values $r_0(t + \Delta t)$, where $\Delta t = 1$ and $\Delta t = 5$ are used as the prediction horizons. Performance in Fig. 4d is quantified by the RMSE between the predicted and true target signals.

## References

1 Lü, J., Yu, X. & Chen, G. Chaos synchronization of general complex dynamical networks. *Physica A: Statistical Mechanics and its Applications* **334**, 281–302 (2004).

2 Cao, H. & Wiersig, J. Dielectric microcavities: Model systems for wave chaos and non-Hermitian physics. *Reviews of Modern Physics* **87**, 61–111 (2015).

3 Bohigas, O., Giannoni, M.-J. & Schmit, C. Characterization of chaotic quantum spectra and universality of level fluctuation laws. *Phys. Rev. Lett.* **52**, 1 (1984).

4 Chialvo, D. R. Emergent complex neural dynamics. *Nature Physics* **6**, 744–750 (2010).

5 Strogatz, S. H. *Nonlinear Dynamics and Chaos: With Applications to Physics, Biology, Chemistry, and Engineering*. (Chapman and Hall/CRC, 2024).

6 Aronson, D. G., Ermentrout, G. B. & Kopell, N. Amplitude response of coupled oscillators. *Physica D: Nonlinear Phenomena* **41**, 403–449 (1990).

7 Koseska, A., Volkov, E. & Kurths, J. Oscillation quenching mechanisms: Amplitude vs. oscillation death. *Physics Reports* **531**, 173–199 (2013).

8 Pikovsky, A. & Rosenblum, M. Synchronization. *Scholarpedia* **2**, 1459 (2007).

9 Sompolinsky, H., Crisanti, A. & Sommers, H.-J. Chaos in random neural networks. *Phys. Rev. Lett.* **61**, 259 (1988).

10 Van Vreeswijk, C. & Sompolinsky, H. Chaos in neuronal networks with balanced excitatory and inhibitory activity. *Science* **274**, 1724–1726 (1996).

11 Rabinovich, M. I., Varona, P., Selverston, A. I. & Abarbanel, H. D. Dynamical principles in neuroscience. *Reviews of Modern Physics* **78**, 1213–1265 (2006).

12 Kadmon, J. & Sompolinsky, H. Transition to chaos in random neuronal networks. *Physical Review X* **5**, 041030 (2015).

13 Preskill, J. Quantum computing in the NISQ era and beyond. *Quantum* **2**, 79 (2018).

14 Arute, F. *et al.* Quantum supremacy using a programmable superconducting processor. *Nature* **574**, 505–510 (2019).

15 Yan, M., Huang, C., Bienstman, P., Tino, P., Lin, W. & Sun, J. Emerging opportunities and challenges for the future of reservoir computing. *Nat. Commun.* **15**, 2056 (2024).

16 Ascoli, A. *et al.* Edge of Chaos Theory Unveils the First and Simplest Ever Reported Hodgkin–Huxley Neuristor. *Adv. Elec. Mater.* **11**, 2400789 (2025).

17 Chua, L. O. Memristors on 'edge of chaos'. *Nature Reviews Electrical Engineering* **1**, 614–627 (2024).

18 Rafayelyan, M., Dong, J., Tan, Y., Krzakala, F. & Gigan, S. Large-scale optical reservoir computing for spatiotemporal chaotic systems prediction. *Phys. Rev. X* **10**, 041037 (2020).

19 Brückerhoff-Plückelmann, F. *et al.* Probabilistic photonic computing with chaotic light. *Nat. Commun.* **15**, 10445 (2024).

20 Appeltant, L. *et al.* Information processing using a single dynamical node as complex system. *Nature Communications* **2**, 468 (2011).

21 Kamimaki, A. *et al.* Chaos in spin-torque oscillator with feedback circuit. *Physical Review Research* **3**, 043216 (2021).

22 Watts, D. J. & Strogatz, S. H. Collective dynamics of 'small-world' networks. *Nature* **393**, 440–442 (1998).

23 Barabási, A.-L. & Bonabeau, E. Scale-free networks. *Scientific american* **288**, 60-69 (2003).

24 Barabási, A.-L. *Network science*. (Cambridge University Press, 2016).

25 Liu, Y.-Y., Slotine, J.-J. & Barabási, A.-L. Controllability of complex networks. *Nature* **473**,

167–173 (2011).

26 Yuan, Z., Zhao, C., Di, Z., Wang, W.-X. & Lai, Y.-C. Exact controllability of complex networks. *Nature Communications* **4**, 2447 (2013).

27 Holme, P. & Saramäki, J. Temporal networks. *Physics Reports* **519**, 97–125 (2012).

28 Millán, A. P. *et al.* Topology shapes dynamics of higher-order networks. *Nature Physics* **21**, 353–361 (2025).

29 Shao, Y., Dahmen, D., Recanatesi, S., Shea-Brown, E. & Ostojic, S. Impact of local connectivity patterns on excitatory-inhibitory network dynamics. *PRX Life* **3**, 023008 (2025).

30 Aljadeff, J., Stern, M. & Sharpee, T. Transition to chaos in random networks with cell-type-specific connectivity. *Physical Review Letters* **114**, 088101 (2015).

31 Kitayama, K.-i. Guiding principle of reservoir computing based on "small-world" network. *Scientific Reports* **12**, 16697 (2022).

32 Kawai, Y., Park, J. & Asada, M. A small-world topology enhances the echo state property and signal propagation in reservoir computing. *Neural Networks* **112**, 15–23 (2019).

33 Strata, P. & Harvey, R. Dale's principle. *Brain Research Bulletin* **50**, 349–350 (1999).

34 Kandel, E. R. & Spencer, W. A. Cellular neurophysiological approaches in the study of learning. *Physiological Reviews* **48**, 65–134 (1968).

35 Byun, S.-S. & Forrester, P. J. *Progress on the Study of the Ginibre Ensembles*. (Springer Nature, 2025).

36 Clark, D. G. & Abbott, L. Theory of coupled neuronal-synaptic dynamics. *Physical Review X* **14**, 021001 (2024).

37 Metz, F. L. Dynamical mean-field theory of complex systems on sparse directed networks.

*Physical Review Letters* **134**, 037401 (2025).

38 Erdős, P. & Rényi, A. On random graphs. Publicationes Mathematicae (Debrecen), Vol. 6. *Publicationes Mathematicae* **6**, 290–297 (1959).

39 Song, H. F. & Wang, X.-J. Simple, distance-dependent formulation of the Watts-Strogatz model for directed and undirected small-world networks. *Physical Review E* **90**, 062801 (2014).

40 Wolf, A., Swift, J. B., Swinney, H. L. & Vastano, J. A. Determining Lyapunov exponents from a time series. *Physica D: Nonlinear Phenomena* **16**, 285–317 (1985).

41 Maslov, S. & Sneppen, K. Specificity and stability in topology of protein networks. *Science* **296**, 910-913 (2002).

42 Bertschinger, N., Natschläger, T. & Legenstein, R. At the edge of chaos: Real-time computations and self-organized criticality in recurrent neural networks. *Advances in Neural Information Processing Systems* **17** (2004).

43 Schuecker, J., Goedeke, S. & Helias, M. Optimal sequence memory in driven random networks. *Physical Review X* **8**, 041029 (2018).

44 Asabuki, T. & Clopath, C. Taming the chaos gently: a predictive alignment learning rule in recurrent neural networks. *Nature Communications* **16**, 6784 (2025).

45 Kesgin, B. U. & Teğin, U. Implementing the analogous neural network using chaotic strange attractors. *Communications Engineering* **3**, 99 (2024).

46 Jaeger, H. The “echo state” approach to analysing and training recurrent neural networks. Report No. GMD Report 148, 13 (German National Research Center for Information Technology, 2001).

47 Poole, B., Lahiri, S., Raghu, M., Sohl-Dickstein, J. & Ganguli, S. Exponential expressivity

in deep neural networks through transient chaos. *Advances in neural information processing systems* **29** (2016).

48 Raghu, M., Poole, B., Kleinberg, J., Ganguli, S. & Sohl-Dickstein, J. in *Proceedings of the 34th International Conference on Machine Learning.* 2847-2854 (PMLR).

49 LeCun, Y., Bottou, L., Bengio, Y. & Haffner, P. Gradient-based learning applied to document recognition. *Proceedings of the IEEE* **86**, 2278–2324 (2002).

50 Chattopadhyay, A., Hassanzadeh, P. & Subramanian, D. Data-driven predictions of a multiscale Lorenz 96 chaotic system using machine-learning methods: reservoir computing, artificial neural network, and long short-term memory network. *Nonlinear Processes in Geophysics* **27**, 373-389 (2020).

51 Jaeger, H. Short term memory in echo state networks. Report No. GMD Report 152, (German National Research Center for Information Technology, Sankt Augustin, Germany, 2001).

52 Dambre, J., Verstraeten, D., Schrauwen, B. & Massar, S. Information processing capacity of dynamical systems. *Scientific Reports* **2**, 514 (2012).

53 Srinivasan, K., Plenz, D. & Girvan, M. Boosting reservoir computing with brain-inspired adaptive control of EI balance. *Nature Communications* **16**, 10212 (2025).

54 Dalgaty, T. *et al.* Mosaic: in-memory computing and routing for small-world spike-based neuromorphic systems. *Nature Communications* **15**, 142 (2024).

55 Davies, M. *et al.* Loihi: A neuromorphic manycore processor with on-chip learning. *IEEE Micro* **38**, 82–99 (2018).

56 Merolla, P. A. *et al.* A million spiking-neuron integrated circuit with a scalable communication network and interface. *Science* **345**, 668–673 (2014).

57 Yu, S., Piao, X. & Park, N. Neuromorphic functions of light in parity-time-symmetric systems. *Adv. Sci.* **6**, 1900771 (2019).

58 Bogaerts, W. *et al.* Programmable photonic circuits. *Nature* **586**, 207-216 (2020).

59 Yu, S. Evolving scattering networks for engineering disorder. *Nat. Comput. Sci.* **3**, 128-138 (2023).

60 Kidger, P. *On Neural Differential Equations*, Oxford, (2021).

61 Fagiolo, G. Clustering in complex directed networks. *Physical Review E—Statistical, Nonlinear, and Soft Matter Physics* **76**, 026107 (2007).

62 Hagberg, A. A., Schult, D. A. & Swart, P. J. in *7th Python in Science Conference (SciPy2008).* (eds Gäel Varoquaux, Travis Vaught, & Jarrod Millman) 11–15.

63 Lorenz, E. in *Seminar on Predictability.* 1–18.

## Data availability

Data used in the current study are available from the corresponding authors upon request and can also be obtained by running the shared codes at 10.5281/zenodo.20280818 in the Zenodo.

## Code availability

Codes used in the current study are available at 10.5281/zenodo.20280818 in the Zenodo.

## Acknowledgements

We acknowledge financial support from the National Research Foundation of Korea (NRF) through the Innovation Research Center (No. RS-2024-00413957), Core Research Grants (No. RS-2026-25469085), Pilot and Feasibility Grants (No. RS-2025-19912971), and Basic Research

Laboratory (No. RS-2024-00397664), all funded by the Korean government. This work was supported by Creative-Pioneering Researchers Program and the BK21 FOUR program of the Education and Research Program for Future ICT Pioneers in 2026, through Seoul National University. We also acknowledge an administrative support from SOFT foundry institute.

## Author Contributions

J.K., N.P., and S.Y. conceived the idea. J.K. and K.K. developed the theoretical tool and performed the numerical analysis. J.K., K.K., and K.P. examined the theoretical and numerical analysis. S.Y. and N.P. supervised the findings of this work. All authors discussed the results and wrote the final manuscript.

## Competing Interests

The authors have no conflicts of interest to declare.

## Additional information

**Correspondence and requests for materials** should be addressed to N.P. or S.Y.

## Figure Legends

**Figure 1. Network topology on neuronal dynamics. a,** A directed neural circuit network with node-wise positive (black) or negative (orange) edges. The initial values are uniformly sampled from the interval [–5, 5]. The node states evolve dynamically over $t$. **b,** Illustration of Dale-type sign constraints, where individual neurons provide strictly excitatory or inhibitory connections. **c-e,** Dynamics in the regular (**c**) small-world (**d**), and random (**e**) networks for $\beta = 10^{-4}$, 0.1, and 0.8, respectively. $N = 300$, $p_0 = 0.07$, and $E = 0.2$ in **c-e**.

**Figure 2. Topological classification of circuit dynamics. a,** Network parameters: clustering coefficient $C$ and average path length $L$, as functions of $\beta$. **b,** Temporal metrics: memory timescale $\tau_c$ and first-passage time $\tau_p$, as functions of $\beta$. $\varepsilon = 0.01$ in estimating $\tau_p$. **c,d,** Time autocorrelation $Q(\varsigma)$ (**c**) and perturbation spreading (**d**) for regular (top; $\beta = 10^{-4}$), small-world (middle; $\beta = 0.1$), and random (bottom; $\beta = 0.8$) networks. The $y$-axis of each subpanel in **d** depicts nodes sorted by their shortest-path graphical distance from the batch of initially perturbed nodes $B$. Dashed lines in **a** and **b** denote $\beta$ values for the subpanels in **c** and **d**. In analysing dynamics, we generate an ensemble of 24 random realizations for each $\beta$ value. All metrics in **a** and **b** are normalized by their maximum values. All other parameters are the same as those in Figs. 1c-1e.

**Figure 3. Chaos-latency phase diagrams**. Phase diagrams are shown over the ($\beta$, $E$) parameter space for the memory metric $\tau_c$ (**a**), latency metric $\tau_p$ (**b**), and the LLE (**c**). The parameters $\beta$ and $E$ values are swept over the ranges [$10^{-4}$, $10^{0}$] and [0.0, 1.0], respectively, with a resolution of 60 × 60. Each point in the parameter space represents an average over an ensemble of 24 realizations. Red dashed lines in **a** and **b** denote $E = 0.2$. **d,** Schematic for edge rewiring. The green and red edges indicate the selected rewired edge pairs. **e,** Phase transition of dynamics induced by edge rewiring, which corresponds to the purple arrow in **a**: starting from $\beta = 10^{-1.15}$ (total 6015 edges) and performing 180 swaps at $t = 200$, then returning to the original state at $t = 400$. All other parameters are the same as those in Fig. 2.

**Figure 4. Topologically programmable chaos for task-specific computation**. **a,** Schematic for reservoir-computing framework based on our neural circuit, applied to three representative tasks: MNIST image classification (left), remote signal tracking (centre), and temporal signal prediction

(right). **b-d,** Task performances as a function of $\beta$ at $E$ = 0.2, spanning regular, small-world, and random network regimes: classification error on MNIST (**b**), root-mean-square error (RMSE) for remote signal tracking (**c**), and RMSE for chaotic time-series prediction at $\Delta t$ = 1 (red) and $\Delta t$ = 5 (blue) (**d**). Each point in **b-d** denotes one random realization of a neural circuit.

# Supplementary Information for "Task-specific programming of chaos in neural circuits"

Jungyoon Kim[1§], Kyuho Kim[1§], Kunwoo Park[1], Namkyoo Park[2†], and Sunkyu Yu[1*]

[1]Intelligent Wave Systems Laboratory, Department of Electrical and Computer Engineering, Seoul National University, Seoul 08826, Korea

[2]Photonic Systems Laboratory, Department of Electrical and Computer Engineering, Seoul National University, Seoul 08826, Korea

E-mail address for correspondence: [†]nkpark@snu.ac.kr, [*]sunkyu.yu@snu.ac.kr

**Note S1. Robustness of the passage time to $\varepsilon$**

The definition of $\tau_p$ in Eq. (2) of the main text requires a threshold parameter $\varepsilon$, which is set to $\varepsilon = 0.01$ in the main analysis. To assess the sensitivity of our results to this hyperparameter, we recomputed the dependence of $\tau_p$ on the rewiring probability $\beta$ over a range of $\varepsilon$ values (Fig. S1). The results show that the qualitative trend is preserved over a broad range of $\varepsilon$, demonstrating the robustness of the first-passage time $\tau_p$.

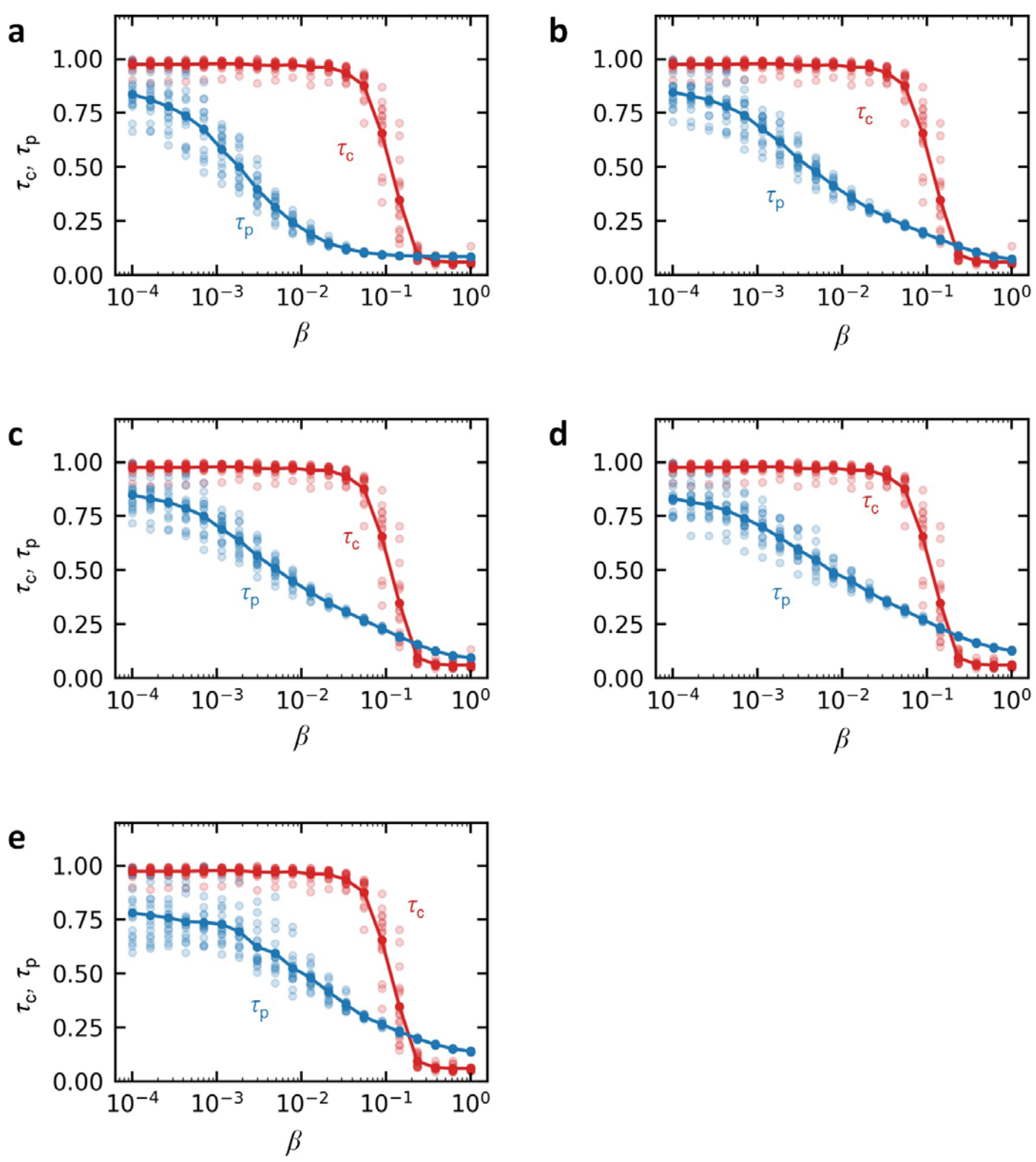

**Fig. S1. Robustness to $\varepsilon$.** $\tau_p$ for $10^{-3}$ (**a**), 0.03 (**b**), 0.05 (**c**), 0.1 (**d**), and 0.2 (**e**). $\tau_c$ is also plotted for comparison.

**Note S2. Size dependency of phase diagrams**

To assess whether the dynamical phase diagrams reported in Fig. 3 of the main text are universal against finite-size effects, we examine their dependence on the network size $N$. Figures S2 and S3 show the memory and latency phase diagrams, respectively, for $N = 100$, 200, 300, 400, and 500. For small networks, the phase diagrams exhibit noticeable finite-size fluctuations, reflecting the stronger influence of individual nodes and links on the collective reservoir dynamics. As $N$ increases, however, the global structures of both phase diagrams become progressively stabilized.

In particular, the qualitative locations and shapes of the observed dynamical regimes become nearly unchanged for $N \geq 300$. Therefore, the system size $N = 300$, used in the main text, is sufficiently large to capture the converged phase-diagram landscape while remaining computationally tractable. These results confirm that the observed dynamical regimes are not finite-size artifacts but represent robust topology- and parameter-dependent behaviours of the reservoir network.

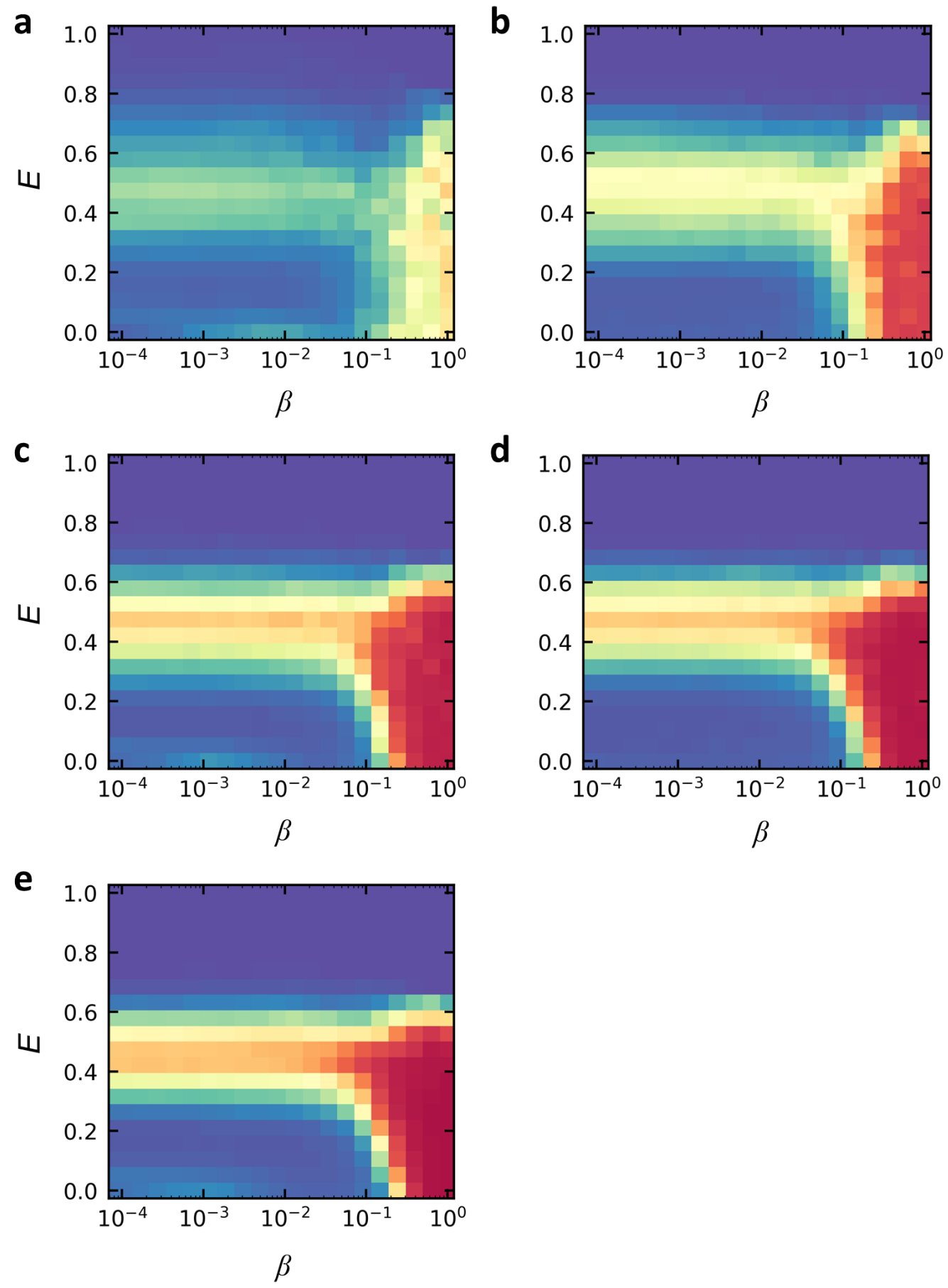

**Fig. S2. Size-dependent memory phase diagram.** Phase diagrams with resolution 20 × 20, for $N$ = 100 (**a**), 200 (**b**), 300 (**c**), 400 (**d**), and 500 (**e**).

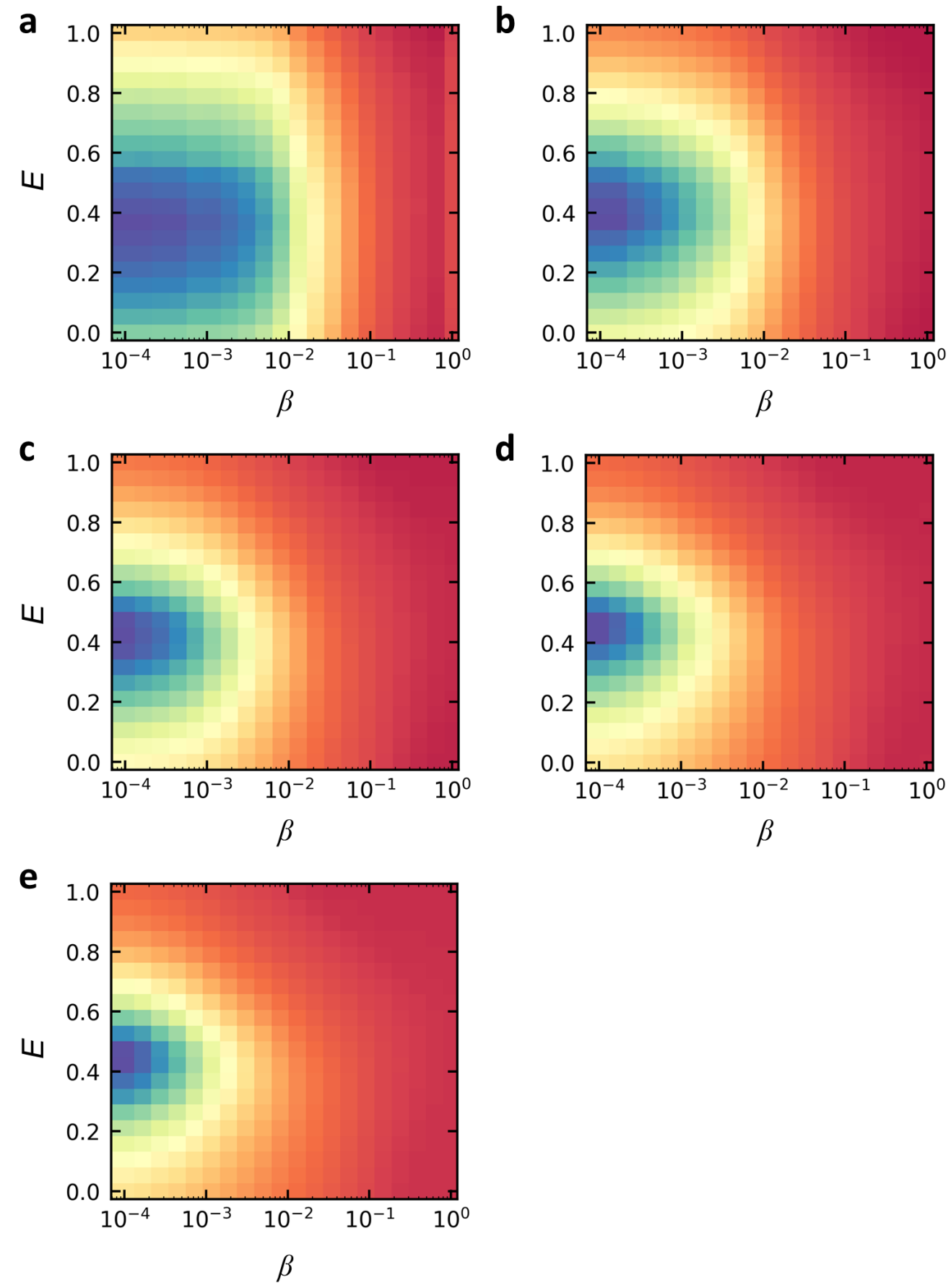

**Fig. S3. Size-dependent latency phase diagram.** Phase diagrams with resolution 20 × 20, for $N$ = 100 (**a**), 200 (**b**), 300 (**c**), 400 (**d**), and 500 (**e**).

**Note S3. Network expressivity**

The expressivity of the echo state network (ESN) is quantified using a geometric measure adapted from prior work[1], which analyses the evolution of low-dimensional input manifolds through a neural network. In this framework, expressivity is associated with the complexity of the embedding of an input manifold into the hidden representation space of the network. We adapt this approach to our ESN by injecting a static input vector $s_i(\theta)$ parameterized by a scalar variable $\theta$, evolving the network dynamics for a fixed amount of transient duration ($t_{\text{transient}} = 10$), and then, analysing the Grassmannian length $\mathcal{L}_G$ of the evolved network $x_i(\theta)$ as a function of $\theta$. For the tangent vector of the trajectory, $v(\theta) = \partial_\theta x(\theta)$, and its normalized form, $\hat{v}(\theta) = v(\theta)/ \|v(\theta)\|$, the Grassmannian length is defined as

$$\mathcal{L}_G = \int \left\| \frac{\partial \hat{v}(\theta)}{\partial \theta} \right\| d\theta \,. \tag{1}$$

In the phase diagram shown in Fig. S4a, we find that the chaotic phases identified in Fig. 3 exhibit substantially higher expressivity than the regular and ordered phases. This relationship is further clarified in Fig. S4b, which shows that increasing $\tau_c$ leads to lower $\mathcal{L}_G$. Consequently, particularly low expressivity is observed in the high -$\beta$ and high-$E$ regime. These results suggest that network topology also governs the expressivity of dynamical neural networks.

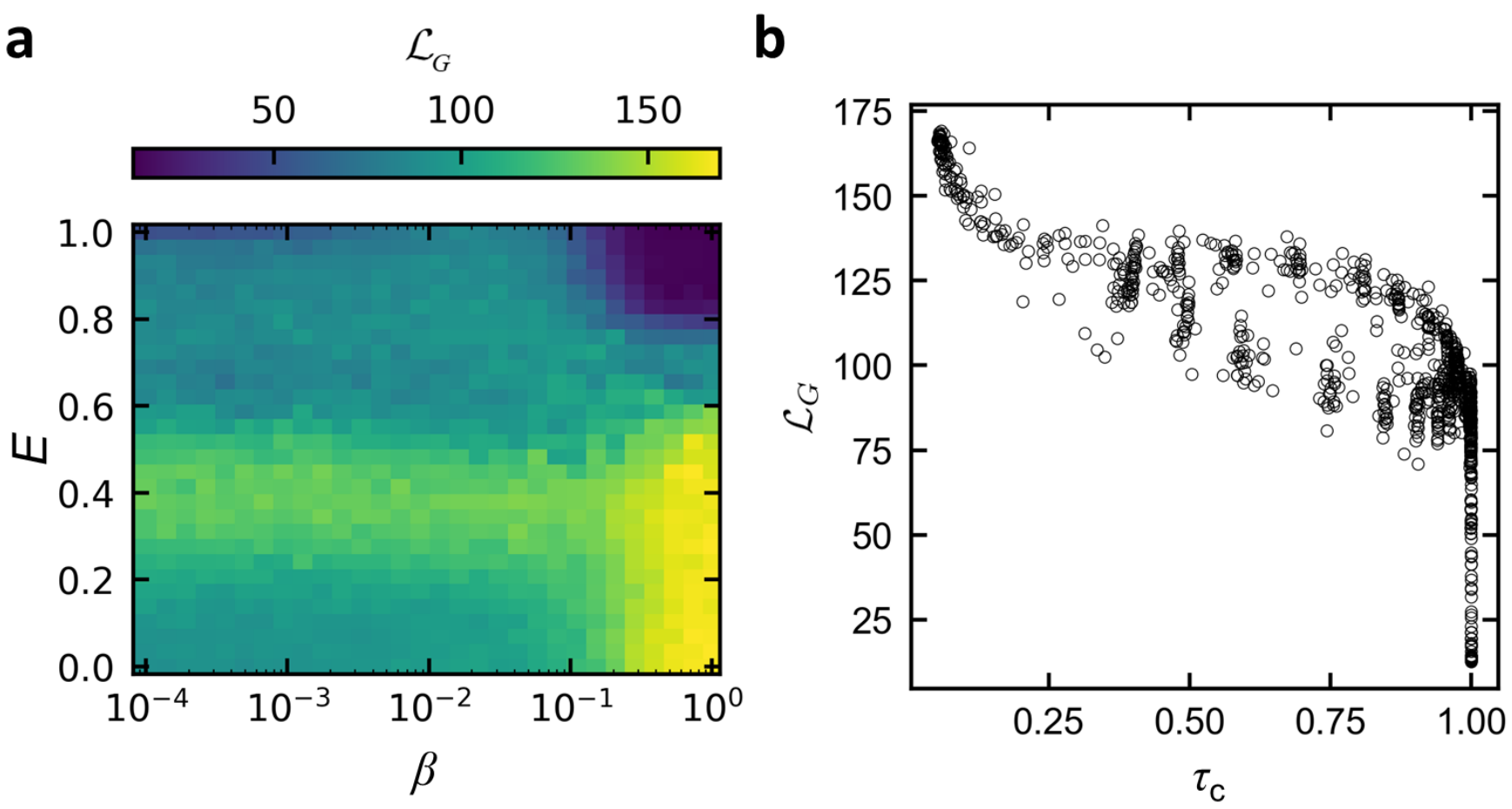

**Fig. S4. Expressivity analysis. a,** The phase diagram of the Grassmannian length $\mathcal{L}_G$, with resolution 30 × 30. **b,** Scatter plot of $\mathcal{L}_G$ versus the dynamical correlation time $\tau_c$.

**References**

1 Poole, B., Lahiri, S., Raghu, M., Sohl-Dickstein, J. & Ganguli, S. Exponential expressivity in deep neural networks through transient chaos. *Advances in neural information processing systems* **29** (2016).